# Beyond Quantities: Machine Learning-based Characterization of Inequality in Infrastructure Quality Provision in Cities

Bo Li [a*], Ali Mostafavi [a]

[a] UrbanResilience.AI Lab, Zachry Department of Civil and Environmental Engineering, Texas A&M University, College Station, TX, 77843, USA

**Abstract**

The objective of this study is to characterize inequality in infrastructure quality across urban areas. While a growing of body of literature has recognized the importance of characterizing infrastructure inequality in cities and provided quantified metrics to inform urban development plans, the majority of the existing approaches focus primarily on measuring the quantity of infrastructure, assuming that more infrastructure is better. Also, the existing research focuses primarily on index-based approaches in which the status of infrastructure provision in urban areas is determined based on assumed subjective weights. The focus on infrastructure quantity and use of indices obtained from subjective weights has hindered the ability to properly examine infrastructure inequality as it pertains to urban inequality and environmental justice considerations. Recognizing this gap, this study examines inequality in infrastructure quality in which infrastructure quality is defined based on the extent to which infrastructure features shape environmental hazard exposure (i.e., air pollution and urban heat) of urban areas. We propose a machine learning-based approach in which infrastructure features that shape environmental hazard exposure are identified and we use the weights obtained by the model to calculate an infrastructure quality provision for spatial areas of cities and accordingly, quantify the extent of inequality in infrastructure quality. The implementation of the model in five metropolitan areas in the U.S. demonstrates the capability of the proposed approach in characterizing inequality in infrastructure quality and capturing city-specific differences in the weights of infrastructure features. The results also show that areas in which low-income populations reside have lower infrastructure quality provision, suggesting the lower infrastructure quality provision as a determinant of urban disparities. Accordingly, the proposed approach can be effectively used to inform integrated urban design strategies to promote infrastructure equity and environmental justice based on data-driven and machine intelligence-based insights.

**Introduction**

The year 2020 witnessed the moment when in the first time in history urban population surpassed rural population; furthermore, the total number of people living in cities is estimated to boom by 150% by 2050 (United Nations, 2020). As cities bear the majority of global population, reducing urban inequality is becoming an important component of realizing the Sustainable Development Goals (SGD) proposed by the United Nations to reduce inequalities (United Nations, 2023). In characterizing urban inequality as a multi-faceted phenomenon, researchers recognize that the distribution of infrastructure systems, in addition to economic factors, is a major source of urban inequality (Nelson, Warnier, & Verma, 2023). Infrastructure systems serve as the backbone of modern cities, as these systems provide essential physical facilities and public services. Inadequate access to infrastructures may cause hardship in multiple dimensions, such as travel inconvenience (Aman & Smith-Colin, 2020; Jiao & Dillivan, 2013), negative health outcomes (Diez Roux & Mair, 2010), community safety threats (Hwang, Joh, & Woo, 2017), as well as social segregation and isolation (Feitosa, Le, & Vlek, 2011; Niembro, Guevara, & Cavanagh, 2021). Moreover, infrastructure systems can persist for a relatively long time, which are even possible to pass the impacts between generations (Grant, 2010; Pandey, Brelsford, & Seto, 2022). The widespread and long-term impacts of infrastructure necessitate the investigation of infrastructure inequality at the city scale.

While a growing body of literature has recognized the importance of characterizing infrastructure inequality in cities and provided quantified metrics to inform urban development plans, current approaches for examining infrastructure inequality have primary limitations that need to be addressed. First, almost all the related studies evaluate infrastructure provision merely based on the quantities of infrastructure, implying the studies share the same premise that utilities created by infrastructure are cumulative with no upper bounds. In other words, more infrastructure is always better. Nevertheless, it is not always the case that more infrastructure brings more utilities, as excessive infrastructures can cause negative effects as well. For instance, roads are often accompanied by noise and air pollution if the traffic volume is high, and the side effects may have disproportionate effects on vulnerable populations (Yildirim & Arefi, 2021). Infrastructure provision inequality evaluation based on quantities can be misleading due to the failure of considering the associations between infrastructures and environment, urban health and sustainability. Besides, infrastructure quantities in certain regions are constrained by land area, which makes it impossible for infrastructure quantity to increase infinitely. Therefore, instead of focusing on the quantities of infrastructure provision, attention should be switched to infrastructure quality. In other words, the ultimate goal of developing urban infrastructures should not be chasing the highest level of infrastructure quantities, but to yield ultimate benefits related to sustainability and health, so as to improve wellbeing without aggravating negative impacts.

Second, most existing studies examine condition and provision variation for only individual infrastructure types, such as tree canopy distribution (Lin, Wang, & Li, 2021; Volin et al., 2020), transportation network (X. Liu, Dai, & Derudder, 2017), and electricity access (Sarkodie & Adams, 2020; Stock, 2021). Such efforts gleaned empirical evidence on disparities of certain infrastructure provision, while assessment on the overall conditions of multiple infrastructure types is still scarce. In a few studies tackling the problem (e.g. (Li, Wang, Zarazaga, Smith-Colin, & Minsker, 2022), overall infrastructure conditions are simply aggregated from each individual type, and equal weights (in most cases subjective) are assumed for all infrastructure types, thus failing to consider the contribution of different infrastructure in shaping urban phenomena such as environmental justice and urban health. Hence, the indicator-based method for quantifying infrastructure provisions based on equal or subjective weights is oversimplified and does not capture the relationships among various infrastructure features and urban sustainability outcomes. These limitations have hindered integrated urban design strategies from simultaneously capturing the intersection of infrastructure development and environmental justice.

Recognizing these important gaps, this study proposes a novel approach for quantifying and assessing infrastructure provision in cities based on a data-driven and machine learning method. This study introduces infrastructure quality (IQ) provision as a new concept, which evaluates the extent to which infrastructure features shape environmental hazard exposures in cities. The concept of infrastructure quality provision recognizes that certain infrastructure features may aggravate the environmental burden while providing convenience. The upshot of a primary focus on quantity of infrastructure as a positive utility, while neglecting the extent to which infrastructure features shape environmental hazards (i.e., air pollution and urban heat) across different areas of cities for developing infrastructure have been misleading, which led to infrastructure and urban development patterns that exacerbate environmental justice, health and sustainability. Thus, departing from the existing approaches, infrastructure quality provision is proposed in this study to better capture the intersection of infrastructure development and environmental justice in measuring and characterizing infrastructure inequality in cities. The core premise of infrastructure quality provision is to maintain an infrastructure provision level at which urban services and economic activities are improved without incurring high environmental hazard risk. Although urban services and economic activities improve with infrastructure quantity, increasing quantities beyond a certain extent could be associated with high environmental hazards. Thus, the trade-off effects of infrastructure provision imply the existence of certain infrastructure provision thresholds at which service and economic activity are optimal while the associated environmental hazard exposure is mild.

To specify and compute infrastructure quality provision, we take advantage of an interpretable machine learning technique to capture the extent to which various infrastructure features and their complex non-linear interactions shape environmental hazard exposure (i.e., air pollution and urban heat). The predictability performance of the models indicates the extent of association between infrastructure features and environmental hazard exposures. In this way, this model could identify whether an optimal threshold exist for each infrastructure feature and if so, the threshold value, which makes it possible for us to calculate infrastructure quality provision quantitively. Fig. 1a compares the core idea of quantity- and quality-based infrastructure provision by schematic representation. Fig.1b shows the data-driven framework developed in this study. The framework consists of three primary components: 1) perform binary classification model to examine how infrastructure features shape environmental hazards and use Shapley Additive explanation (SHAP) method to determine the magnitude and direction of each infrastructure component towards environmental hazards; 2) compute overall infrastructure quality provision based on the machine learning-based weights and thresholds; 3) compute infrastructure quality provision inequality spatially and across income levels. Each component is discussed in more detail in the following subsections.

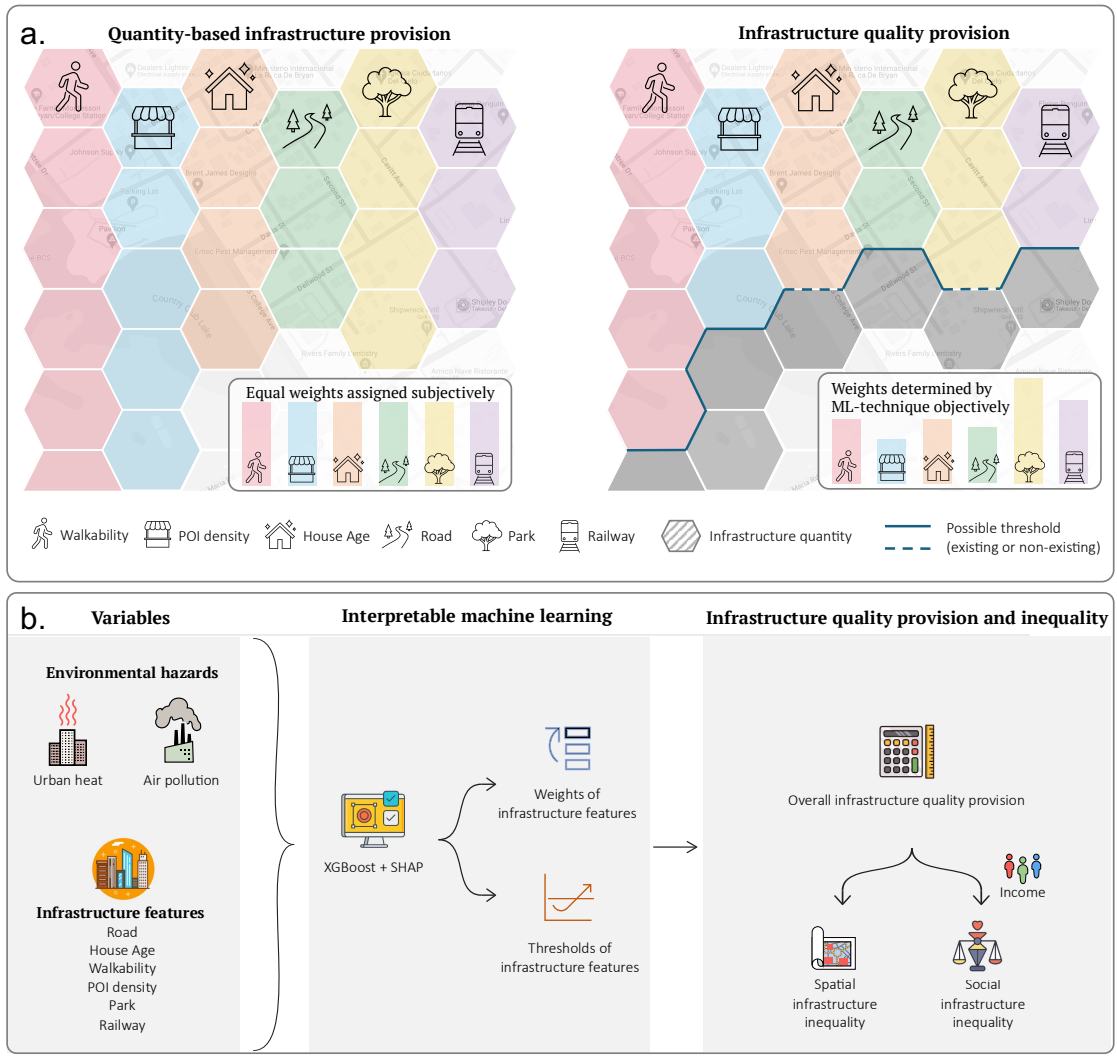

**Fig 1.** Conceptual illustration of study approach. **a.** Schematic representation of infrastructure quality provision and comparison with quantity-based infrastructure provision. Quantity-based infrastructure provision measurement assumes more infrastructure is better and assign equal weights across infrastructure features. In contrast, infrastructure quality provision proposes that thresholds regarding infrastructure quantities may exist, which, when exceeded, may be associated with greater environmental hazards. The weights among infrastructure features are determined by feature importance retrieved from an interpretable machine learning method. **b.** Data and methods. Two environmental hazards, urban heat and air pollution, are considered to examine the extent to which infrastructure features shape hazard exposure in specifying infrastructure quality provision. The XGBoost model and the SHAP method are used to obtain weights and thresholds for infrastructure features. Then overall infrastructure quality provision is calculated to reveal potential spatial and social infrastructure inequality.

Furthermore, the interpretable machine learning method could also reveal relative infrastructure feature importance in shaping environmental hazards, which enables us to consider the contribution level of each infrastructure feature to calculate infrastructure quality provision. Using feature importance as a proxy for the weights of infrastructure feature, we construct an integrated infrastructure quality provision metric for each geographical unit. We evaluate infrastructure quality provision index at the census tract level in five cities of the United States: Houston, Dallas, Los Angeles, Detroit, and Chicago. Further the study reveals both spatial and socio-demographic inequality in infrastructure quality provision.

The main novel aspects of this study are fivefold: (1) this study introduces infrastructure quality provision as an important alternative to the mostly quantity-based infrastructure metrics used for evaluation of infrastructure inequality in cities; (2) this study specifies infrastructure quality provision based on the extent to which infrastructure features shape environmental hazard exposures for each city, and hence, facilitating the city-specific evaluation of the contribution of different features on infrastructure quality provision; (3) this study leveraged data-driven metrics and machine intelligence in capturing the complex and non-linear interactions among infrastructure features at the intersection of infrastructure development and environmental justice while examining infrastructure inequality; (4) departing from rather index-based methods based on subjective weights for specifying infrastructure provision scores, this study instead harnesses machine intelligence in determining city-specific weights for each infrastructure feature; (5) this study unveils both spatial and socio-economic inequality in infrastructure quality provision across areas of cities. These contributions offer significant advancements and move the interdisciplinary fields of urban computing, infrastructure, sustainability and geography closer to more data-driven and machine intelligence-based characterization of infrastructure inequality. These contributions inform integrated urban design strategies to better evaluate infrastructure development patterns by shifting focus from quantity to quality of infrastructure provision at the intersection with environmental justice issues.

**Results**

**Contributions of various infrastructure features to environmental hazard**

Contributions of infrastructure features in shaping environmental risk may not be constant across features and cities. Here, we used a machine learning model (ML) to examine the extent to which infrastructure features explain variability in the extent of environmental hazard (i.e., air pollution and urban heat) across different areas of cities. After evaluating different ML approaches, we found XGBoost to provide the best performance. The prediction performance statistics are displayed in Method section. Based on the ML classification model, feature importance analysis was performed using SHAP. SHAP analysis reveals detailed information on the extent to which infrastructure features shape environmental hazards across different areas of cities. For multiple infrastructure features across the five study areas, Fig. 2 provides an overview of relative feature importance in predicting environmental hazard category. Each bar displays the mean absolute SHAP value across the dataset, representing the average contribution of certain infrastructure feature towards high environmental hazard risk. Bats are stacked in decreasing order, which shows the ranking of features importance. The rankings vary greatly across cities. For example, the feature importance of rail ranks first in the city of Chicago, while was at the bottom of the list in Dallas, Los Angeles, and Detroit. The variability of feature importance across cities indicates that a consistent pattern of relative importance does not exist, and hence the current index-based methods fail to capture differences across cities. Thus, it is essential to use ML-based techniques, such as the one reported in this paper, to specify weights of various infrastructure features in calculating infrastructure provision scores of spatial areas in order to capture city-specific contexts.

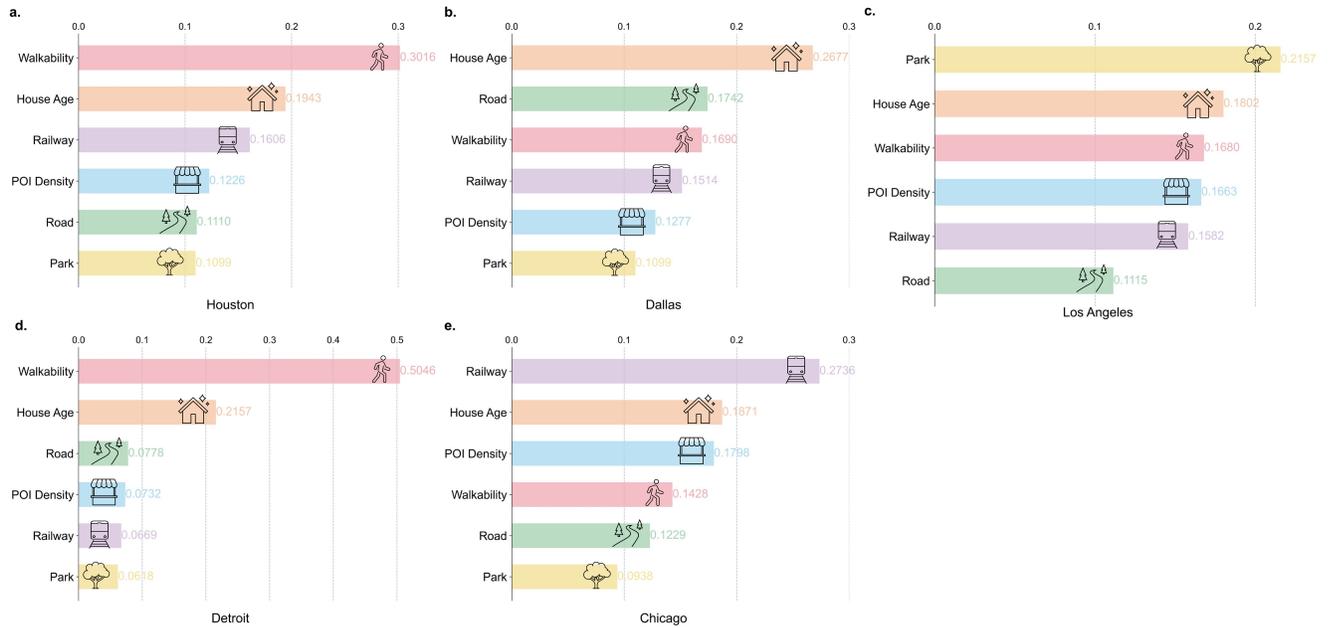

**Fig. 2** Weights of infrastructure features for the five U.S. cities. Weights are determined based on global importance of infrastructure features and are normalized using the softmax function. **a.** Houston; **b.** Dallas; **c.** Los Angeles; **d.** Detroit; **e.** Chicago. The rankings of weights vary greatly across cities.

In addition to specifying the relative importance of infrastructure features, we further analyzed the SHAP dependence plots to identify the threshold value at which an infrastructure feature value is considered to have the optimum utility for infrastructure quality provision. On the SHAP dependence scatter plot, each dot represents an instance. The x-axis of dots shows feature value, and the y-axis shows the corresponding SHAP value for each instance. Thus, SHAP dependence plots display the effects of a single infrastructure feature on the environmental hazard. We fit the scatter plots with local weighted scatterplots smoothing (LOWESS) technique to capture patterns. LOWESS is a non-parametric regression technique, which fits a weighted regression line to localized subsets of the data (Cleveland, 1979). The fitted curve could reflect the overall trend of the data while keep flexibility in capturing local variations. The shadow displays a 95% confidence interval of the fitted lines. The fitted curves reveal important patterns regarding the optimum feature value for each infrastructure feature. Figs. 3a and 3b present fitted curves intersecting the x-axis and indicating an increasing tendency of SHAP values to change from negative to positive as the infrastructure quantity increases. The shape of the curve illustrates that the contribution towards high environmental hazard increases with infrastructure quantity. Since positive SHAP values indicate a positive relationship of infrastructure features towards high environmental hazard, and vice versa, the intersection point of the curve with x-axis signifies the inflection point where the direction of feature contribution changes. For example, in Fig.3a, 10.68% is an inflection point for railway, beyond which railway and environmental hazard risk has a positive association. This means census tracts with more than 10.68% of tract area within a 1-mile buffer of a railway will face greater environmental hazards. Thus, the intersection point can serve as a threshold where adverse effects of more infrastructure start to emerge. Fig. 3c and 3d present an opposite pattern: the overall tendency of the fitted curve is decreasing, indicating that more infrastructure quantity is associated with a lower environmental hazard exposure. Similarly, the third pattern shows that the fitted curve remains negative, meaning that the infrastructure retains a negative association with high environmental hazard. In both cases, infrastructure growth is not associated with high environmental hazard, so there would not exist a threshold signifying the optimal level of infrastructure provision. Comparing the complete results (available in supplementary document) for the six infrastructure features across five cities, we identified the thresholds consistently exist for the same features across cities. For example, for road and railway, quantities exceeding the optimal level may cause high environmental hazard. This finding coincides with the previous studies. For example, Mukherjee et al. (2020) and Ferm and Sjöberg (2015) observed significant PM2.5 increments in near-road areas due to roadway emissions. Heavy road traffic and paved roads also account for worsening the urban heat issue (Akbari & Kolokotsa, 2016; Mohajerani, Bakaric, & Jeffrey-Bailey, 2017; Tzavali, Paravantis, Mihalakakou, Fotiadi, & Stigka, 2015). Thresholds are not identified in park and house age, which means higher provision of parks and larger proportion of newly built houses are associated with low environmental hazard, which might be due to the ability of park to mitigate urban heat island effects (Algretawee, Rayburg, & Neave, 2019; Yao et al., 2022) and reduce air pollution (Kumar et al., 2019). Also, newer homes tend to be constructed in newly developed areas with land development provisions that alleviate environmental hazards such as urban heat. The results for the point-of-interest (POI) density (Figs. 3e and 3f) and walkability features vary across cities, which could be due to the differences in city structure, public transportation distribution, and facility centralization. These two infrastructure features need to be analyzed on a city-by-city basis.

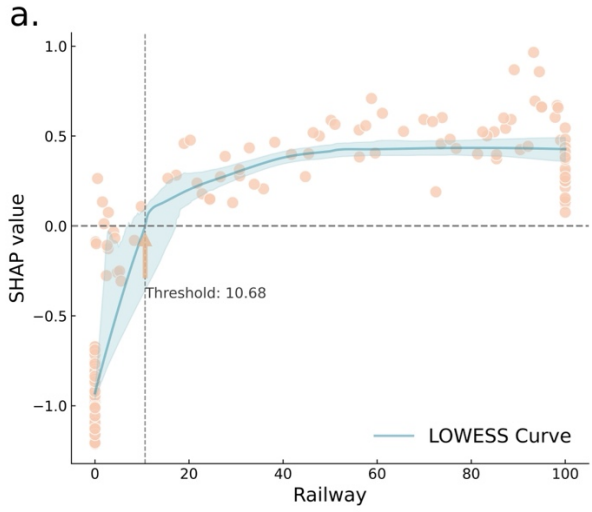
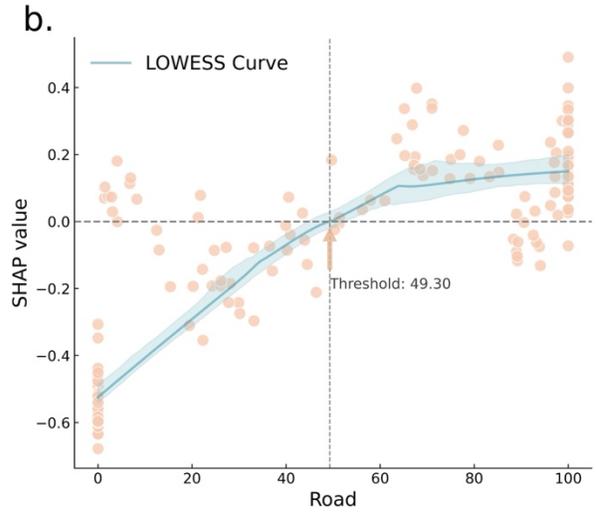
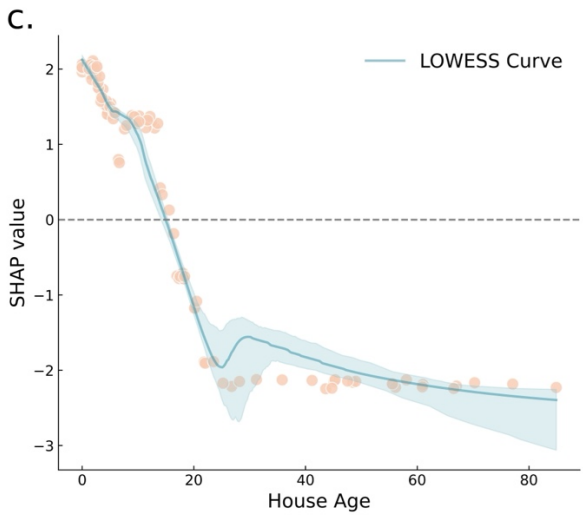
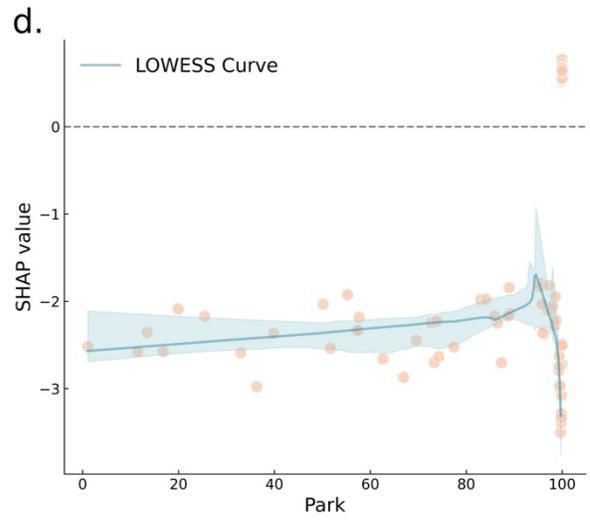
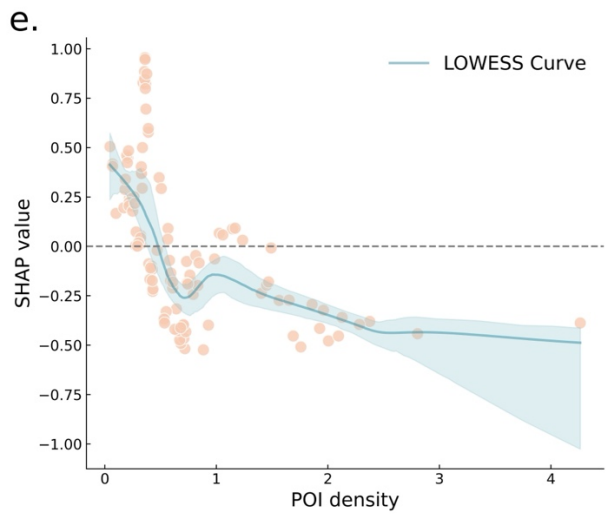
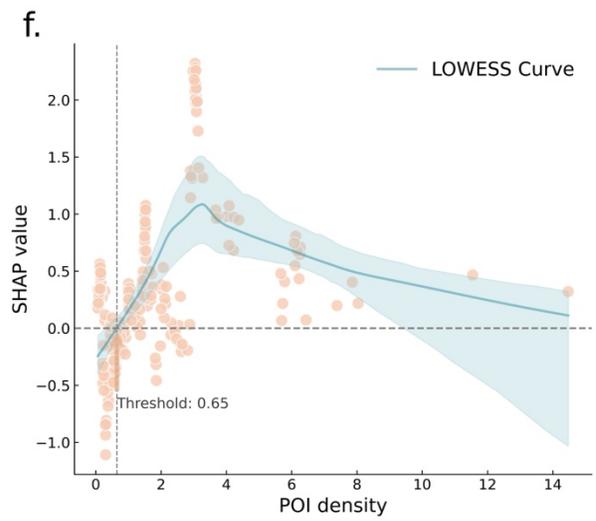

**Fig. 3** SHAP dependence plot. This plot displays the directions and extents to which infrastructure features shape environmental hazards. Possible thresholds can be identified from the plot. Greater quantity of road and railway contribute to greater environmental hazards, while greater quantity of parks and newer houses contribute to lower environmental hazards. **a.** Rail in Houston; **b.** Road in Houston; **c.** House age in Detroit; **d.** Park in Los Angeles; **e.** POI density in Dallas; **f.** POI density in Chicago

**Infrastructure quality provision distribution at the census tract level**

One of the most significant distinctions of this research is to examine infrastructure provision based on both quality and quantity. We acknowledge the conventional view that more infrastructure is conducive to improve wellbeing, while the growth of infrastructure should not result in excessive environmental burden. Using the previously discussed results related to the associations of between each infrastructure feature and environmental hazard, as well as the relative importance of infrastructure in shaping environmental hazard, we computed infrastructure provision index at census tract level for the studied five cities. Further details on computing method are available in Method section. Fig. 4 visualizes the spatial distribution of infrastructure quality provision for five cities in the United States.

The results reveal that infrastructure quality provision level is not equally distributed across census tracts of each city. To measure the extent of inequality in infrastructure quality provision, we implemented the infrastructure inequality index (Pandey et al., 2022) and displayed the result in Table 1. The index is designed to range from 0 to 1, where 0 represents no inequality, and 1 represents maximum inequality. All the studied cities have a status of intermediate inequality, with Dallas presenting the highest level of infrastructure quality provision inequality, and Los Angeles, the lowest. Juxtaposing Dallas versus Los Angeles, the extent of inequality in infrastructure quality in Dallas is 27% greater than that of Los Angeles. These results show the utility of the proposed infrastructure quality provision metric and method for inter- and intra-city comparisons.

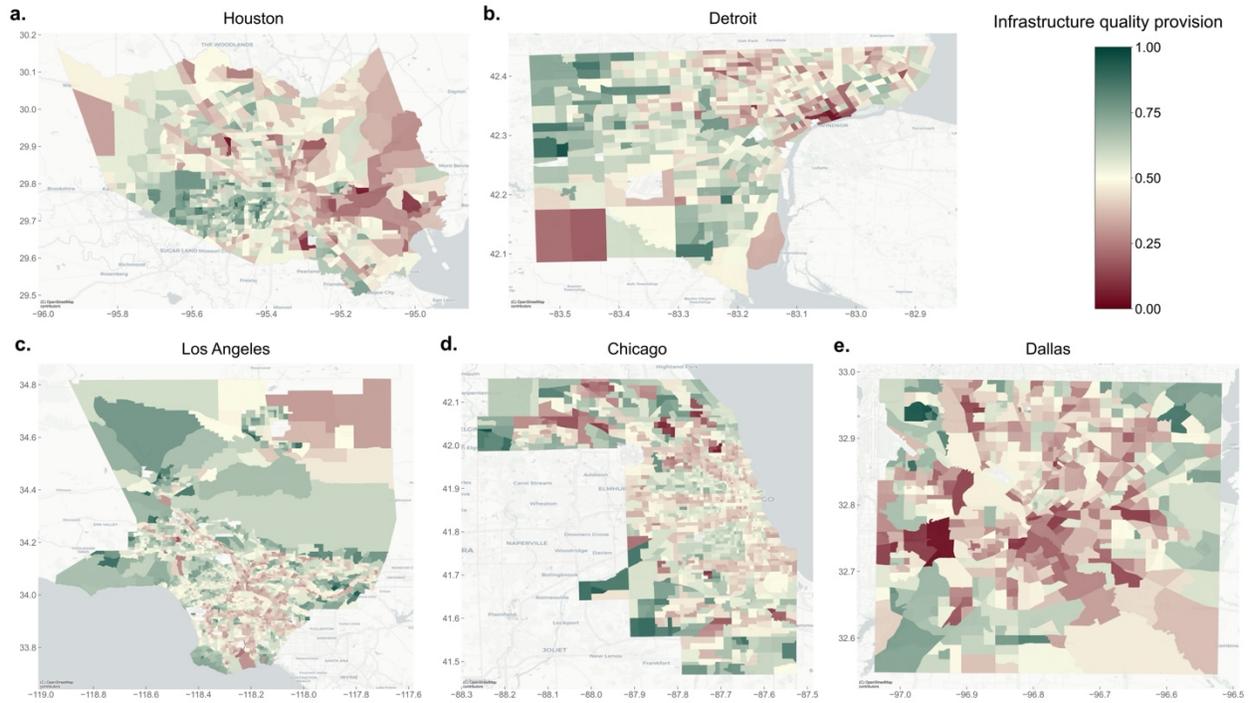

**Fig. 4** Geographical distribution of infrastructure quality provision across five cities. Infrastructure quality provision ranges from 0 to 1, with 0 representing worst infrastructure provision. **a.** Houston; **b.** Detroit; **c.** Los Angeles; **d.** Chicago; **e.** Dallas.

**Table 1.** Infrastructure inequality index

| Area | Houston | Dallas | Los Angeles | Detroit | Chicago |
|---|---|---|---|---|---|
| Infrastructure inequality index | 0.316 | 0.352 | 0.277 | 0.282 | 0.309 |

**Comparisons between quality- and quantity-based infrastructure provisions**

For comparison purposes, we calculated a quantity-based infrastructure provision based on the same infrastructure feature data for the five cities. We grouped census tracts within each city into five groups according to the percentile of quantity-based and quality-based infrastructure provision score. Fig. 5 displays the results for Los Angeles (Results for other cities are available in Supplementary Materials). The spatial distributions of quantity-based and quality-based infrastructure provision (Fig. 5a and 5b) present significant differences for most areas. Census tracts with top 20% infrastructure quality provision rank at the bottom 20% category in terms of quantity-based infrastructure provision and vice versa. Fig.5c displays the distribution of median income across census tracts in Los Angeles, which shows a similar spatial pattern to that of infrastructure quality provision. The bar plots in Figs. 5d and 5e confirm the pattern that census tracts with higher median income are associated with higher infrastructure quality provision level, while the association between income and the quantity-based infrastructure provision shows an inverse relationship.

The comparison between the quantity-based and quality-based infrastructure provisions has an important implication. The ultimate goal of infrastructure provision is to promote well-being, environmental justice, and economic prosperity. Examining infrastructure provision based on quantity and subjective weights mis-characterizes infrastructure provision in different areas of cities yielding better provision scores in areas with more infrastructure and having greater environmental hazards. Also, quantity-based infrastructure provision shows better infrastructure provision in low-income areas sufferings from greater environmental hazards and environmental injustice issues. Hence, the existing quantity-based and subjective weights assigning approaches lead to mis-characterizing infrastructure provision, which fails to promote infrastructure development strategies to enhance environmental justice and equality.

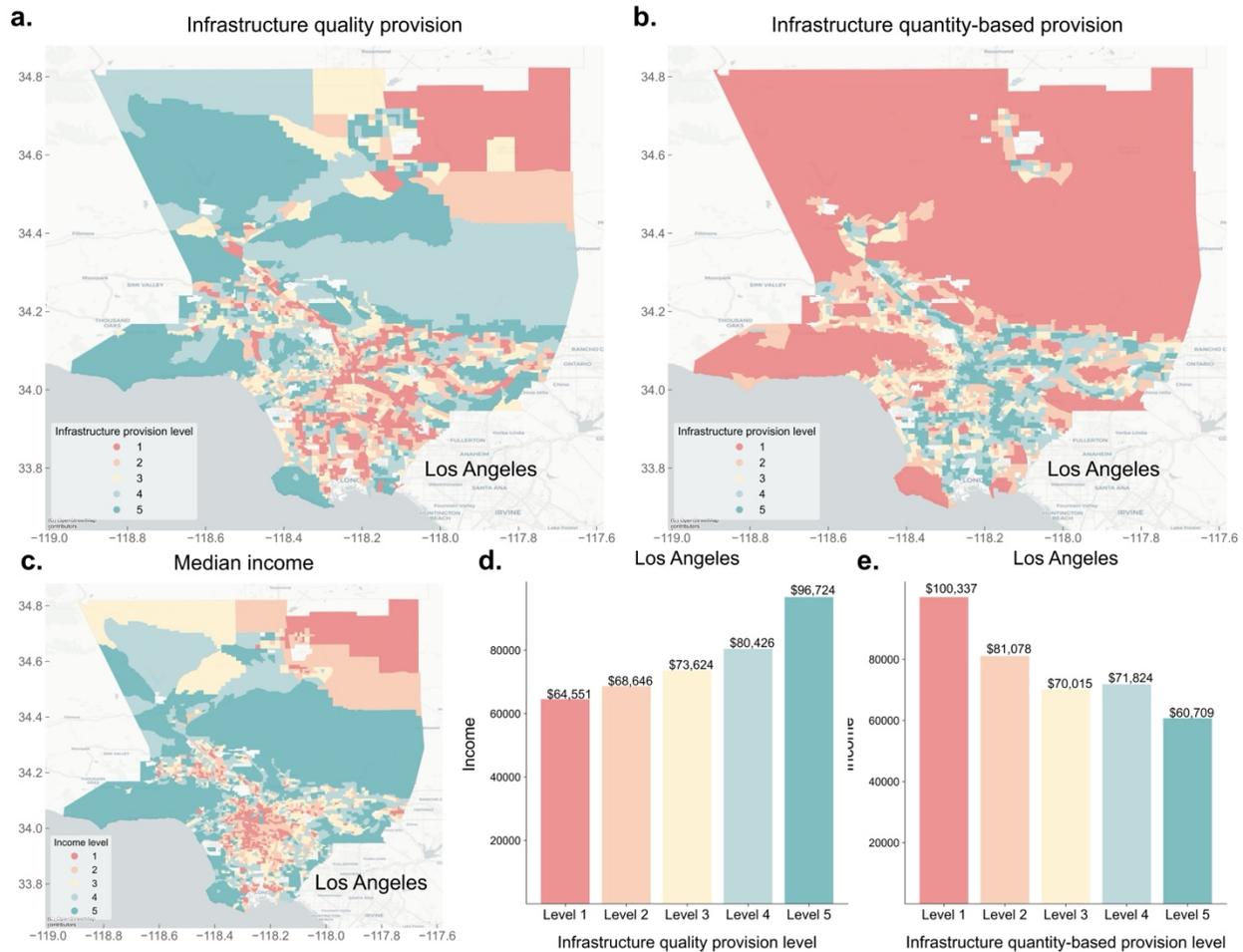

**Fig. 5** Comparisons between quantity- and quality-based infrastructure provision in Los Angeles. **a.** Spatial distribution of quality-based infrastructure provision; **b.** Spatial distribution of quantity-based infrastructure provision; for each subplot, census tracts are categorized into five levels using quantiles. Levels 1 through 5 denotes increasing infrastructure provision. In general, the distribution for quantity- and quality-based infrastructure provision present reversed patterns of distribution. **c.** Spatial distribution of income with income increasing from level 1 through level 5. The pattern of income distribution is similar to that of infrastructure quality provision. **d.** Average income for each quality-based infrastructure provision level; **e.** Average income for each quantity-based infrastructure provision level.

**Social disparities on infrastructure quality provision**

In the next steps, we examined infrastructure quality provision disparity across income groups. Using the median value of infrastructure quality provision values, we split census tracts of each city into two groups: a better provisioned group and a worse provisioned group. Results (Fig. 6) reveal significant disparities between infrastructure quality provision groups in terms of income.

Census tracts in the better provisioned group have a higher median income, consistent across all five cities. For example, the median income of census tracts with better infrastructure provision is 27% higher than the ones with worse infrastructure provision in Los Angeles, which is the largest gap among the five cities. Interestingly, Los Angeles has the least level of spatial infrastructure inequality, yet it has the greatest infrastructure inequality across income groups. This result depicts the significance of measuring and evaluating inequality in infrastructure quality both in terms of spatial and social disparities. We also divided census tracts within a city into low- and high-income groups divided at the $50^{th}$ percentile of census tract median income. Fig. 7 presents the distribution of infrastructure quality provision in low- vs. high- income groups. A consistent pattern emerges for the differences between two income groups in Dallas, Los Angeles, and Detroit: high-income groups have higher levels of infrastructure quality provision. For example, approximately 20% high income census tracts exhibit infrastructure quality provision lower than 0.4, while in low-income group the proportion is about 40%. This finding reveals the different patterns of socio-demographic inequality in infrastructure quality provision across different cities. In addition, the area between the two curves (low-income versus. high-income curves) in Fig. 7 provides a quantified value for the extent of social inequality in infrastructure quality provision in cities for cross-city comparison. As shown in Fig. 7, Dallas and Detroit have the greatest social inequality in infrastructure quality provision, followed by Los Angeles. Houston, while having the second greatest spatial inequality, has the lowest social inequality in infrastructure quality provision. These results highlight the importance of concurrent evaluation of spatial and social inequalities in infrastructure quality for inter- and intra-city comparison.

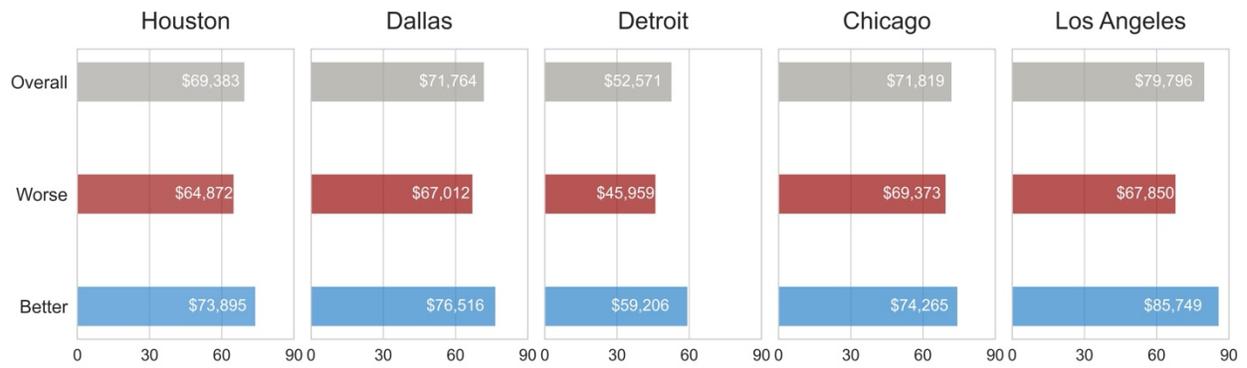

**Fig. 6** Average income of different infrastructure quality provision groups. Census tracts in each city are divided into two groups by infrastructure provision scores. Results for the five cities present the consistent pattern that areas with better infrastructure quality provision have higher income compared those with worst infrastructure quality provision.

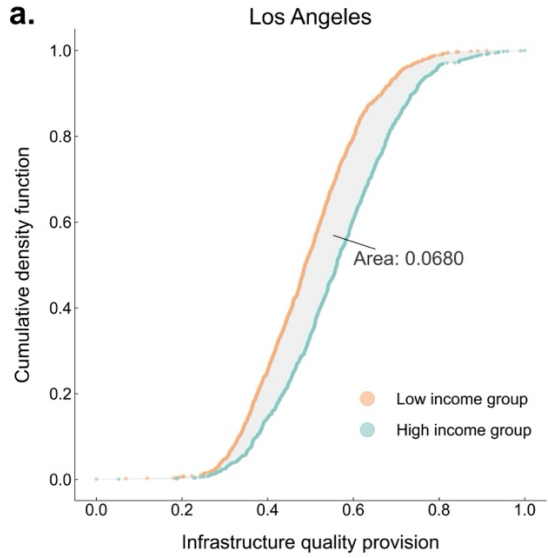
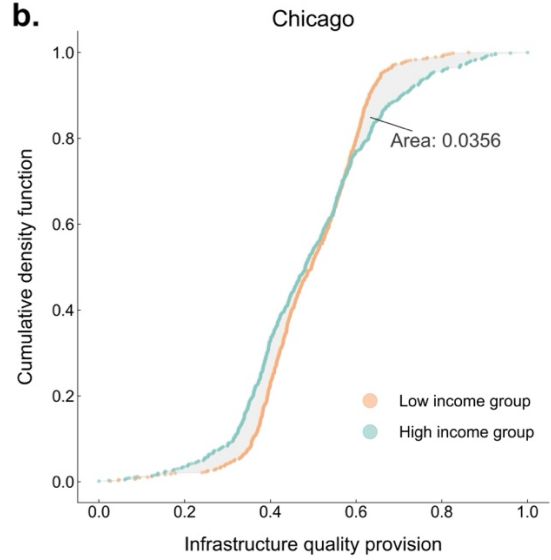
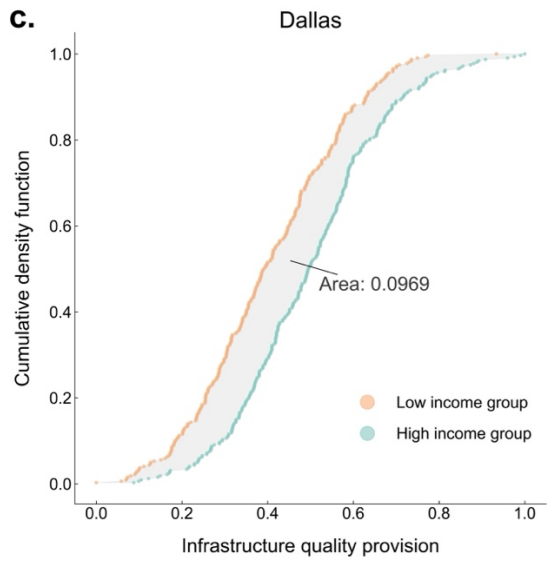
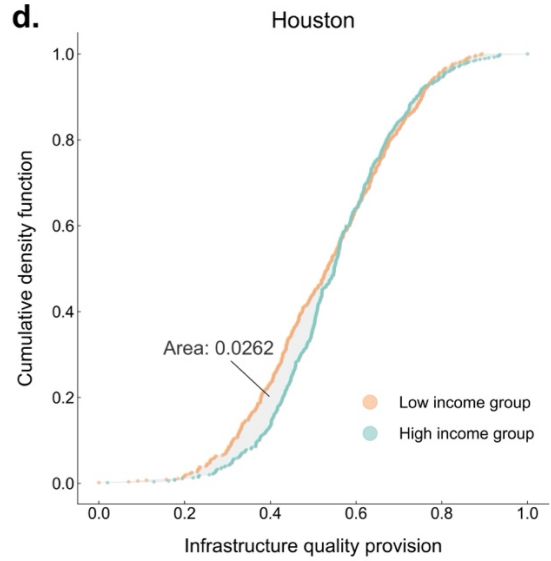
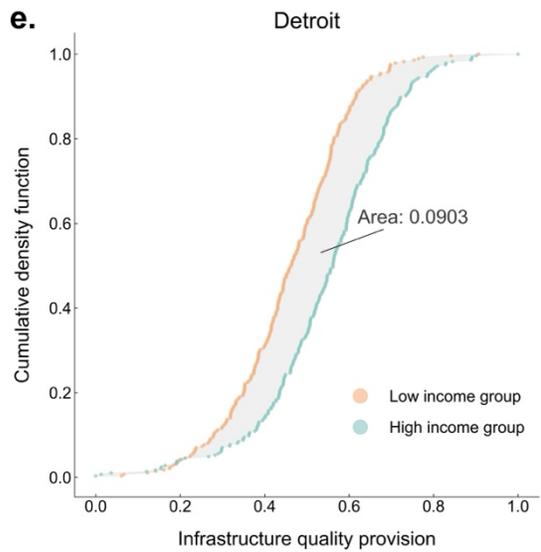

**Fig. 7** Distribution of infrastructure quality provision in low-and high-income groups. The curves are cumulative density function of infrastructure quality provision. Significant differences are observed between low-income and high-income curves. Areas between the curves quantifies the disparity extent between two income groups. **a.** Los Angeles; **b.** Chicago; **c.** Dallas; **d.** Houston; **e.** Detroit. Dallas has the greatest extent of disparity while Houston has the least.

**Discussion**

This study presents a data-driven and ML-based framework for assessing urban infrastructure quality provision and its inequality for inter- and intra-city analyses. The main novel aspects of this study are fivefold: First, this study introduces infrastructure quality provision as an important alternative to the mostly quantity-based infrastructure metrics used for evaluation of infrastructure inequality in cities; The main premise of infrastructure quality provision is that infrastructure development should not merely pursue quantity of growth as the ultimate goal, but take sustainability, environmental justice and economic prosperity into consideration. This study compared quantity-based infrastructure provision with the quality-based approach within the same cities. Significant mischaracterization was revealed that areas with better quantity-based infrastructure provision having greater environmental hazards in which lower-income residents reside. The result indicates that shifting from quantity to quality in terms of infrastructure provision assessment provides a new and more effective approach to integrating environmental justice issues into infrastructure development plans and projects.

Second, the approach presented in this study enables specification of infrastructure quality provision based on the extent to which infrastructure features shape environmental hazard exposures (i.e. urban heat and air quality) for each city, and hence, facilitated the city-specific evaluation of the contribution of different features to infrastructure quality provision. The results identified the extent to which each infrastructure feature contributes to high environmental hazard risk. The results demonstrated that road and railway present similar patterns that if the infrastructure provision level exceeds certain thresholds, a positive association between the infrastructure and environmental hazard risk will emerge. The identification of the threshold indicates that the prevalent assumption that more infrastructure is better is flawed, in that excessive infrastructure could exacerbate environmental risk burden. However, this pattern does not apply to all the infrastructure feature types. For example, greater quantity of parks consistently contribute to alleviating environmental hazard, which conforms with the opinion of Gill, Handley, Ennos, and Pauleit (2007) and O. Y. Liu and Russo (2021). The contributions of infrastructure features such as POI density and walkability, to the infrastructure quality provision vary among different cities, which might be due to differences in characteristics. For example, walkability is a composite indicator which reflects built environment characteristics, such as diversity of land use, transit stops, and road intersection density. The subtle variations of these factors can lead to the differences in the association between infrastructure features and environmental hazard risk. The findings show the capability of the proposed approach in specifying city-specific infrastructure quality provision.

Third, this study leveraged data-driven metrics and machine intelligence in capturing the complex and non-linear interactions among infrastructure features at the intersection of infrastructure development and environmental justice while examining infrastructure inequality, which may not be fully captured if using conventional statistics models (Fan, Xu, Natarajan, & Mostafavi, 2023). This study first applied the XGBoost model to specify the associations between infrastructure features and environmental hazards, then applied the SHAP method to interpret the contributions of infrastructure features in shaping environmental hazard. The interpretable machine learning framework implemented in this study can be generalized to deepen the understanding subtle relationships in an urban context.

Fourth, departing from index-based methods based on subjective weights for specifying infrastructure provision scores, this study instead harnessed machine intelligence to determine city-specific weights for each infrastructure feature. This study provides a solution to address the weight assignment issue among multiple infrastructure components regarding comprehensive infrastructure system provision. We objectively assign weights to various infrastructure components based on the overall magnitude of the infrastructure feature importance towards shaping environmental hazard risk. Compared to previous practices that assigned equal weight or determined weights subjectively, this weighting system captures the dynamics between multiple infrastructures and environmental hazards, and also reduces subjectivity. This procedure could benefit urban planners and policymakers by enabling them to tailor priorities for developing infrastructures for each urban area.

Finally, based on quality provision assessment, this study reveals both spatial and socio-demographic inequality in all studied urban areas. We applied an infrastructure inequality index developed by Pandey et al. (2022); the results show all cities exhibit an intermediate level of infrastructure inequality at census tract level. Among the cities, Dallas has the highest extent of infrastructure quality provision inequality. This study also uncovers the disparity of infrastructure quality provision considering community median income. Communities with better infrastructure quality provision are characterized with higher community median income, the amount of which can be up to 20% more. The distribution indicates that low-income communities have a greater proportion which suffers from low-level of infrastructure quality provision. This finding discloses imbalanced infrastructure development both among spatial areas and income groups, which could inform prioritized investments in disadvantaged communities. The focus on infrastructure quality provides a new way for integrating environmental justice issues into infrastructure development plans and projects. These contributions offer significant advancements and move the interdisciplinary fields of urban computing, infrastructure, sustainability, and geography closer to more data-driven and machine intelligence-based characterization of infrastructure inequality. The contributions of the study also inform integrated urban design strategies to better evaluate infrastructure development patterns by shifting focus from quantity to quality of infrastructure provision at the intersection with environmental justice issues.

This study has some limitations. First, although this study provides a framework to assess overall provision in a multi-component infrastructure system, this study does not cover all the infrastructure types due to data availability issues. The incompleteness may affect the accuracy and thoroughness of infrastructure provision assessment. However, the data-driven and machine learning bases approach can be easily transferrable to other contexts or could include more infrastructure types. Second, this study considers only cases where, at most, one threshold exists for each infrastructure component according to the result we obtained, while multiple thresholds may present. Future work is needed to incorporate the more complex case to extend the applicability of the infrastructure quality provision assessment method.

**Data**

Five cities in the U.S. were studied in this research: Houston, Dallas, Los Angeles, Chicago, and Detroit. The five cities differ in geographical locations, population density, and the stage in urbanization process. The variety of the research objects facilitates examining the generalization capability of our proposed approach to assess infrastructure quality provision.

**Infrastructure features**

For the infrastructure system features, we examined six infrastructure features: high-volume road, railway, house age, recreational parks, walkability and POI density. The selected infrastructure features support various needs of urban residents, covering transportation, living conditions and access to facilities. Apart from POI density, the other five features were retrieved from the Environmental Burden Module (EBM) of Environmental Justice Index (Centers for Disease Control and Prevention and Agency for Toxic Substances Disease Registry, 2022). The module collects features from dependable sources, such as the United States Environmental Protection Agency and Centers for Disease Control and Prevention (CDC) with standardized data collection methods, ensuring the accuracy and reliability of the data (Owusu et al., 2022). All features are computed at the census tract level, which is commonly used as a proxy for neighborhoods in the U.S. For high-volume roads, the feature is the proportion of tract area within a 1-mile buffer of a high-volume street or road. Similarly, railways are represented by the proportion of tract area within a 1-mile buffer of a railway. The feature of house age is the proportion of occupied housing units built prior to 1980. Recreational parks are represented by the proportion of tract area within a 1-mile buffer of a park, recreational area, or public forest. The walkability feature is a comprehensive measure of built environments that affects the probability that residents use walking as a mode of transportation. The Environmental Burden Module adopts National Walkability Index Score (United States Environmental Protection Agency, 2021) to indicate walkability. The index considers street intersection density, proximity to transit stops, and diversity of land uses to determine walkability scores. POI density indicates the distribution and concentration level of various facilities within a given region. We retrieved POI location data from SafeGraph (SafeGraph, 2020), then aggregated POIs at census tract level. Then we divided the total number of POIs of each census tract by area to calculate POI density, expressed as the number of POIs per square kilometer in the census tract.

**Environmental hazards**

This study considered two types of environmental hazard: urban heat and air pollution. Air pollution data, also acquired from the Environmental Burden Module, is indicated by the three-year average number of days above daily standard for PM2.5. We retrieved extreme heat data from National Environmental Public Health Tracking Network published by CDC (Centers for Disease Control and Prevention). Extreme heat is represented by the annually total number of days in which an extreme heat event was ongoing. Extreme heat events refer to the period during which a temperature threshold is surpassed for at least two or three consecutive days. Relative temperature threshold is adopted as the 95$^{th}$ percentile, which makes the definition of extreme heat event free from the impacts of climatic and geographical conditions. This indicator is population-weighted when aggregated from gridded raw data to the census tract level. Thus, this data can be a proxy of urban population heat exposure. Built environment features such as greenspace and urban form can create differentials among geographical units (Hsu, Sheriff, Chakraborty, & Manya, 2021).

**Other data**

To uncover potential socio-demographic inequalities regarding infrastructure quality provision, this study incorporates the 2020 annual median income of census tracts retrieved from American Community Survey (ACS)(U.S. Census Bureau, 2020).

All variables and indicators used in this paper are listed in Table 3 for better clarity.

Table 3. Variable Information

| Feature Name | Indicator | Data Source |
|---|---|---|
| High volume roads | proportion of tract area within 1-mile buffer of a high-volume street or road | EBM |
| Railways | proportion of tract area within 1-mile buffer of a railway | EBM |
| House age | proportion of occupied housing units built prior to 1980 | EBM |
| Recreational parks | proportion of tract area within 1-mi buffer of a park, recreational area, or public forest. | EBM |
| Walkability | National Walkability Index Score | EBM |
| POI density | the number of POIs per square kilometer | SafeGraph |
| Air pollution | three-year average number of days above daily standard for PM2.5 | EBM |
| Extreme heat | the annually total number of days in which an extreme heat event was ongoing | CDC |
| Income | 2020 annual median income of census tracts | ACS |

**Methods**

**XGBoost**

This study performed XGBoost, an ensemble learning method, to classify environmental hazard level based on infrastructure features. XGBoost stands for "Extreme Gradient Boosting", which uses decision trees as weak learner, then combines many weak learners to form a strong learner. XGBoost outperforms other methods in performance and execution speed, thus becoming one of the most popular techniques in data mining challenges (Chen & Guestrin, 2016). As a tree-based model, XGBoost could satisfy the need for both accuracy and interpretability on investigating the relationship between infrastructure provision and environmental hazards. Methods such as regression are easy to understand, while the linear assumption between outcome and predictors sometimes are too strong to be compliant with the actual situation. Relaxing the linearity assumption and utilizing tree-based models can mitigate the biases on estimation (Wu, Jia, Feng, Li, & Kuang, 2023).

Data labelling is the first step to perform binary classification machine learning model. Since commonly acknowledged criteria for determining environmental risk level regarding extreme heat and air pollution are not available, as well as considering the differences between cities, such as climatic and geographic conditions, it is not realistic and reasonable to set one-size-fit-all thresholds for labeling environmental risk. Therefore, we used K-means clustering method to cluster the census tracts within a city according to extreme heat and air pollution indicators. Two environmental hazard risk levels are obtained from clustering. In this way, we captured the relative position (i.e., environmental hazard level) of a census tract within the entire dataset and assigned corresponding labels (0 for low environmental hazard risk, and 1 for high environmental risk). Silhouette score, which measures performance of the clustering is displayed in the supplementary documents. To eliminate the impacts of category imbalance, we adopted Synthetic Minority Oversampling Technique (SMOTE), a method combining over-sampling the minority category and under-sampling the majority category, to achieve better classification performance (Chawla, Bowyer, Hall, & Kegelmeyer, 2002).

Before implementing XGBoost, we first split the dataset for each city with an 80/20 ratio, meaning we used 80% of the data as training data to predict the remaining 20%. To improve the model performance, we applied random search and performed a ten-fold cross validation to tune a series of hyperparameters, including max depth, learning rate, gamma, minimum child weight and estimators. Supplemental Table 2 showed the range values for specific hyperparameters and the optimal values to minimize F1 score. Following the hyperparameter tuning, we trained the model with the training dataset with the optimized hyperparameters, and then tested the model performance on testing dataset.

Results for environmental hazard risk classification based on infrastructure features are listed in Table 4.

Table 4. XGBoost model performance for different cities

| Area | F1 score |
|---|---|
| Houston | 0.7624 |
| Dallas | 0.7773 |
| Los Angeles | 0.8379 |
| Detroit | 0.9310 |

**SHAP method**

Traditional machine learning methods provide only black box structures, which impedes the understanding of why a model makes a certain prediction. The emergence of interpretable machine learning methods has bridged the gap by allowing users to look into the models' internal structures and thus enhance model interpretability. Among the various methods, SHAP (SHapley Additive explanation) has been commonly used because it can be easily applied to various types of machine learning tasks, and be visualized in an intuitive manner (Kim & Lee, 2023; Peng, Zhang, Li, Wang, & Yuan, 2023; Wu et al., 2023). SHAP originates from concept of classic Shapley values, a cooperative game theory-based approach, which fairly distributes contributions of features when they collectively achieve a certain outcome (Shapley, 1953). Shapley value of a feature can be interpreted as the marginal contribution to model prediction averaged over all possible models with different combination of features. Based on that, Lundberg and Lee (2017) proposed SHAP values and provides approximation approaches to overcome challenges of computing SHAP values. Later Lundberg et al. (2020) proposed TreeExplainer, an approach which is able to achieve efficient and exact computation of Shapley values on tree-based machine learning models, such as random forest, and gradient boosted trees. TreeExplainer provides insights on both global and local interpretability. Global interpretability means SHAP is able to capture feature importance by averaging SHAP values of each feature across the entire dataset. Global interpretability helps to tell the magnitude and direction of features towards certain outcomes. Local interpretability analyzes each feature's contribution to outcome of individual observations. Moreover, the location explanation provided by TreeExplainer can also distinguish between main effects and interaction effects of features. Further local explanations can be integrated to capture richer and more accurate information of model's behavior. This study performed SHAP analysis on the correctly classified instances in the testing dataset.

In this study and for determining the weights of each infrastructure feature in computing the infrastructure provision score, we adopted SHAP dependence plots to uncover how each infrastructure feature value impacts environmental hazard of every sample in the testing dataset. We acquire information about the direction of the effects, and whether turning points of the direction exist by observing the SHAP dependence plot. Different infrastructure features may have various magnitudes of effects on environmental hazard. Thus, we took advantage of the global insight provided by SHAP, using feature importance score to quantify the contribution of each infrastructure feature in calculating the infrastructure quality provision score.

**Calculate infrastructure quality provision**

We used SHAP dependence plots to investigate the existence of thresholds for each of the infrastructure features. As a result, two scenarios are possible: one scenario is that a threshold exists. In this case, the threshold can be regarded as the optimal level of the infrastructure provision, since infrastructure provision beyond the threshold will cause high environmental hazard, while below the threshold means the infrastructure level may not sufficiently satisfy needs of residents in terms of service. Another scenario is the threshold does not exist for the infrastructure feature. In this case, we followed the common practice of assuming more infrastructure is better, so the optimal level is the maximum level of the infrastructure. To keep consistency of terms, we also refer to the maximum level of infrastructure in this case as the "threshold". Since thresholds in both scenarios represent the optimal levels of infrastructure features, the distance to the threshold can be a measure to quantify infrastructure quality provision of individual infrastructure types. To assess the overall quality provision of multiple infrastructure components, the weight of each infrastructure feature needs to be determined. We assigned weights using global SHAP values for each infrastructure feature, because the SHAP value represents the contribution to high environmental hazard risk. Thus, overall infrastructure provision for each census tract of a city was calculated by:

$$Infrastructure\ Quality\ Provision = 1 - normalized(\sum_{i=1}^{6} W_i' |X_i - T_i|) \qquad \text{Eq (1)}$$

where $W_i'$ is the normalized weight of infrastructure component $X_i$, and $T_i$ is the threshold value for infrastructure component $X_i$. We adopted the softmax function defined by

$$W_i' = \frac{\exp(W_i)}{\sum_{i=1}^{6} \exp(W_i)} \qquad \text{Eq (2)}$$

to calculate $W_i'$, so that all normalized weights sum to 1.

$W_i'$ is $|X_i - T_i|$ denotes the gap between actual level of infrastructure $X_i$ and the optimal level, multiplied by the weight of $X_i$, which reflects the overall infrastructure conditions of the census tract. Then the weighted sum $\sum_{i=1}^{6} W_i' |X_i - T_i|$ was calculated and normalized by min-max scaling equation as:

$$Y_i' = \frac{Y_i - \min(Y_i)}{\max(Y_i) - \min(Y_i)} \qquad \text{Eq (3)}$$

where $Y_i$ is the original value of the weighted sum $\sum_{i=1}^{6} W_i' |X_i - T_i|$, and $Y_i'$ is the normalized value. Such normalization helps scale $Y_i'$ to the range [0,1]. Larger value of $Y_i'$ means a larger deviation from the current infrastructure level to the optimal level, which indicates worse infrastructure provision. Thus, we used $1 - Y_i'$ to flip the result, so that larger value of infrastructure quality provision corresponds to better infrastructure provision in the census tract.

**Calculate quantity-based infrastructure provision**

To calculate quantity-based infrastructure provision, we followed the assumption that more infrastructure is better, which means the maximum level of infrastructure is the optimal level. Thus, we used the distance to the maximum level of infrastructure feature as the measure to quantify infrastructure quantity-based provision. Further, we assigned equal weights across multiple infrastructure features. The overall quantity-based infrastructure provision is calculated as

$$Infrastructure\ Quantity-based\ Provision \qquad \text{Eq (4)}$$

$$= 1 - normalized(\sum_{i=1}^{6} W_i |X_i - T_i|)$$

where $W_i$ is the equally assigned weight of infrastructure component $X_i$, and $T_i$ is the maximum value for infrastructure component $X_i$ within the city.

**Measure inequality index in infrastructure quality provision**

To measure spatial inequality in infrastructure quality provision, we applied the index developed by Pandey et al. (2022). Compared to other economic inequality measurements such as Gini index or entropy-based index, this measurement considers the statistical distribution properties of infrastructures, including constrained growth, as well as binomial or beta distribution. Thus, this measure is more appropriate for measuring infrastructure inequality. Infrastructure inequality index ($I$) is calculated as

$$I = \frac{\sigma}{\sqrt{u(1-u)}}; 0 < \mu < 1 \qquad \text{Eq (5)}$$

where $\sigma$ is the standard deviation of infrastructure quality provision values across all geographical units in the studied area, and $u$ is the mean of infrastructure quality provision correspondingly. The range of $I$ is between 0 and 1, with greater value meaning higher extent of inequality. We applied this measurement respectively to the studied cities: Houston, Dallas, Los Angeles, Chicago and Detroit with infrastructure quality provision data, so as to address the extent, if any, to infrastructure provision inequality exist.

**Data Availability**

The datasets used in this paper are publicly accessible and cited in this paper.

**Code Availability**

The code that supports the findings of this study is available from the corresponding author upon request.

**Acknowledgements**

This work was supported by the National Science Foundation under Grant CMMI-1846069 (CAREER). Any opinions, findings, conclusions, or recommendations expressed in this research are those of the authors and do not necessarily reflect the view of the funding agencies.

**Author contributions**

**Bo Li**: Conceptualization, Methodology, Software, Formal analysis, Investigation, Writing – original draft, Writing – review & editing, Visualization. **Ali Mostafavi**: Conceptualization, Methodology, Writing—Reviewing and Editing, Supervision, Funding acquisition.

**Competing interests**

The authors declare no competing interests.

**Additional information**

Supplementary material associated with this article can be found in the attached document.

35